# Edge states at the boundary of graphene-like and Lieb lattices

I. V. Kozlov[1] , Yu. A. Kolesnichenko[2]

*B.Verkin Institute for Low Temperature Physics and Engineering of the National Academy of Sciences of Ukraine, 47 Nauky Ave., Kharkiv, 61103, Ukraine*

Properties of the boundary of two conductors in a quantizing magnetic field are studied: with conventional Dirac charge carriers and so-called "pseudospin-1" fermions, which are realized in graphene-like and Lieb lattices respectively. It is shown that edge states arise that relate the properties of conductors to the trivial and nontrivial Berry phase. These edge states lead to the appearance of a characteristic series of root features in the density of states.

## 1. Introduction

Currently, many different topological materials are known [1], which are described by different effective spin values, which determine their general properties. In addition to the conventional Dirac fermions [2], so-called "spin-1" fermions [3], as well as Rarita fermions [4], mathematically described by pseudospin 3/2 attract attention. So-called spin-1 fermions, described by E. Majorana in 1928 [5], have attracted attention only recently. They are characterized by a number of properties typical for chiral fermions, such as the presence of a cone in the energy spectrum, the root dependence of Landau levels on the magnetic field, the Klein paradox [6,7], but they differ in the Berry phase. differ in Berry phase, which is equal to $\pi$ for the conventional Dirac cone, and trivial for "spin-1" fermions. From a theoretical point of view, the simplest implementation of "spin-1" fermions on a lattice should apparently be considered the Lieb lattice, which also attracts interest due to the presence of a "Flat Band" in the energy spectrum and is currently made artificially [8], on optical lattices and in organic conductors [6].

A boundary of two media with different topological order can be characterized by series of properties, the existence of which the existence of which is associated with the difference in the nature of the two materials. A well-known example is the surface conductivity of a topological insulator [9]. Although it is worth mentioning that the anomalous near-surface current arising at near-surface magnetic levels in ordinary conductors was already known and was explained in the theory of the static skin effect in the language of traditional electronic theory long before the advent of the theory of topological materials [10].

---

[1] kozlov@ilt.kharkov.ua
[2] kolesnichenko@ilt.kharkov.ua

In the presented work, a boundary of two different topological conductors will be considered: conductors with charge carriers of the Dirac type and "spin-1" fermions, which are realized in a graphene-like and a Lieb lattice, respectively. The paper is organized as follows. Section 2 describes the model. In Section 3 respectively the properties of the boundary without magnetic field are presented. In Section 4 the behavior of the system in a quantizing magnetic field is described (4.1), the magnetic energy levels of the edge states (4.2) are calculated, the form of the density of states (4.3) is given.

## 2. Model

Let us consider the boundary of a graphene-like and Lieb lattices [11], as shown in Fig. 1. We restrict ourselves to the standard nearest neighbor approximation for the electronic energy spectrum:

$$E = \sum_{i,j} t_{i,j} \bar{\phi}_i \phi_j, \qquad (1)$$

where the overlap integral $t_{i,j} = t_G$ for the graphene-like lattice, $t_{i,j} = t_L$ for the Lieb lattice, $t_{i,j} = t_M$ between atoms of two different types, $\phi_i$ -- a probability amplitude of finding an electron near a site $i$ The "Armchair" geometry of the boundary of the graphene-like lattice with a rotation of the Lieb lattice was chosen to preserve the possibility of electron tunneling between the neighborhoods of the Dirac points of both lattices. In this way, the value of a component along the boundary of the momentum at the conical points of both conductors coincides, which is conserved:

$$\vec{k}_C^I = (+\frac{4\pi\hbar}{3a_G}, 0), \quad \vec{k}_C^{II} = (-\frac{4\pi\hbar}{3a_G}, 0), \quad \vec{k}_C^L = \left(\frac{\sqrt{2}\pi\hbar}{a_L}, 0\right), \qquad (2)$$

where the lattice periods $a_G$ and $a_L$ are shown in Fig. 1 and satisfy the relation $\sqrt{3}a_G = \sqrt{2}a_L = a_B$. Let us restrict ourselves to the case of the same Fermi velocity at both edges, which ensures the coincidence of the classical energy spectrum of the two lattices in the vicinity of the cone points, $E = \pm v_F \left|\vec{p} - \hbar\vec{k}_C^{I,II,L}\right|$. In this way, the influence of classical effects on the barrier is reduced, clearly revealing its quantum and topological nature. This condition is

equivalent to the restriction on the overlap integrals $t_G$ and $t_L$, which determine the Fermi velocities $v_G^F = \sqrt{3} a_G t_G / (2\hbar)$ (the graphene-like lattice) and $v_G^L = a_L t_L / \hbar$ (the Lieb lattice): in accordance with the tight binding approximation.

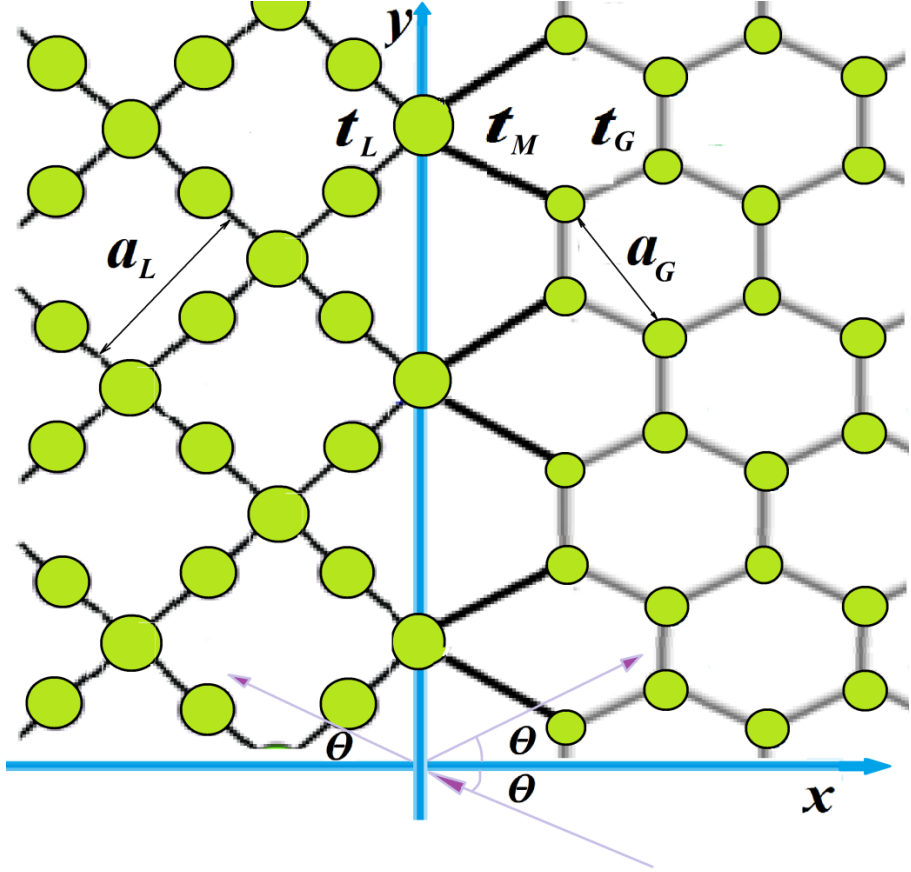

**Fig.1.** The boundary of two lattices and their parameters.

$$t_L = \frac{\sqrt{3}}{2} t_G, \tag{3}$$

but we will assume that the overlap integral $t_M$ between atoms of different lattices can be arbitrary. We will consider the states having a positive energy:

$$0 < E \ll t_{G,L,M}, \tag{4}$$

and also small enough that the deviation from the linearity in the energy spectrum can be neglected, that is equivalent to the smallness of the wave vector $\vec{k}$ measured from the cone point,

$$a_{L,G} k \ll 1, \qquad \vec{p} = \hbar \vec{k}_C^{I,II,L} + \hbar \vec{k}, \tag{5}$$

where $\vec{p}$ -- electron momentum measured from the center of the Brillouin zone.

For the wave function in the graphene we will use the usual spinor notation $\varphi = (\varphi_B, \varphi_A)^T$, where spinor components correspond directly to the values $\phi_i = \varphi_{A,B}(r_i)$ (1) on atoms <u>in</u> $r_i$ with a similar index in Fig. 1. Thus, the energy operator has the form:

$$\hat{\varepsilon} = v_F (\mp \hat{\sigma}_x k_x + \hat{\sigma}_y k_y), \tag{6}$$

where the sign "-" corresponds to the cone point $\vec{p}_C^I$, and the sign "+" -- $\vec{p}_C^{II}$. The eigenfunctions of operator (6) can be written as $\varphi^I(r) = \varphi_{(\pi-\theta)}^G e^{i(\vec{k}_C^I + \vec{k})\vec{r}}$ and $\varphi^{II}(r) = \varphi_\theta^G e^{i(\vec{k}_C^{II} + \vec{k})\vec{r}}$, where

$$\varphi_\theta^G = \frac{1}{\sqrt{2}} \begin{pmatrix} e^{-i\theta/2} \\ e^{+i\theta/2} \end{pmatrix}, \tag{7}$$

the angle $\theta$ determines the direction of the wave vector $\vec{k} = (k\cos\theta, k\sin\theta)$.

Similarly, for the Lieb lattice we write the wave function as $\varphi = (\varphi_1, \varphi_2, \varphi_0)^T$. The energy operator depending on the wave vector $\vec{k} = (k\cos\theta, k\sin\theta)$ has the form:

$$\hat{\varepsilon} \begin{pmatrix} \varphi_1 \\ \varphi_2 \\ \varphi_0 \end{pmatrix} = -i v_F \hbar k \left( \hat{R}_x \cos\left(\frac{\pi}{4} + \theta\right) - \hat{R}_y \sin\left(\frac{\pi}{4} + \theta\right) \right) \begin{pmatrix} \varphi_1 \\ \varphi_2 \\ \varphi_0 \end{pmatrix}, \tag{8}$$

where matrices

$$\hat{R}_x = \begin{pmatrix} 0 & 0 & 1 \\ 0 & 0 & 0 \\ -1 & 0 & 0 \end{pmatrix}, \hat{R}_y = \begin{pmatrix} 0 & 0 & 0 \\ 0 & 0 & -1 \\ 0 & 1 & 0 \end{pmatrix}. \tag{9}$$

Accordingly, the eigenfunctions of the energy operator have the form $\varphi^L(r) = \varphi_\theta^L e^{i(\vec{k}_C^L + \vec{k})\vec{r}}$, where

$$\varphi_\theta^L = \frac{1}{\sqrt{2}} \begin{pmatrix} i\cos(\pi/4+\theta) \\ i\sin(\pi/4+\theta) \\ 1 \end{pmatrix}. \tag{10}$$

## 3. Properties of the boundary without a magnetic field

**3.1.** The energy spectrum of graphene contains two Dirac points $\vec{k}_C^{I,II} = (\pm\frac{4\pi\hbar}{3a_G},0)$, that is, it has additional degeneracy in comparing to the Lieb lattice (2) which has the single point $\vec{k}_C^L = \left(\frac{\sqrt{2}\pi\hbar}{a_L},0\right)$. This leads to different behavior of the wave function depending on its polarization. There will be either partial tunneling of the electron from graphene into the Lieb lattice, or its complete reflection back for polarization of a certain type. It can be said that the boundary of the two lattices in Fig. 1 plays the role of a kind of filter, preventing tunneling for a certain type of polarization. The wave function of the last before reflection has the form:

$$\varphi_{in}^M(\vec{r}) = \sin\frac{\theta}{2}\varphi_\theta^G e^{i(\vec{k}_C^I+\vec{k})\vec{r}} + \cos\frac{\theta}{2}\varphi_{(\pi-\theta)}^G e^{i(\vec{k}_C^{II}+\vec{k})\vec{r}}, \tag{11}$$

where $\vec{k}_{in} = (k\sin\theta, k\cos\theta)$ and the definition (7) is used. The wave function of the reflected from the interface electron can be written as:

$$\varphi_{out}^M(\vec{r}) = -\cos\frac{\theta}{2}\varphi_{(\pi-\theta)}^G e^{i(\vec{k}_C^I+\vec{k}^*)\vec{r}} - \sin\frac{\theta}{2}\varphi_\theta^G e^{i(\vec{k}_C^{II}+\vec{k}^*)\vec{r}}, \tag{12}$$

here the angle $\theta$ corresponds to the expression (11), $\vec{k}^* = (-k\sin\theta, k\cos\theta)$.

For opposite polarization (which differs from (11) by the arguments of the trigonometric functions $\theta \to \theta+\pi$), the boundary of the conductors will be characterized by the transmission coefficient:

$$T = \frac{1}{\frac{1}{8}\frac{t_M^2}{t_G t_L}\left(1+\frac{3}{\cos^2\theta}\right)+\frac{1}{6}\frac{t_G t_L}{t_M^2}\cos^2\theta+\frac{1}{2}}. \tag{13}$$

There is no "supertunneling" for (13), $T < 1$ for all values of the angle $\theta$ and overlap integrals. The maximum value $T_{max} = \left(\frac{1}{\sqrt{3}}+\frac{1}{2}\right)^{-1}$ corresponds to $t_M = t_L$ and $\theta = 0$. In papers [3,12] authors claimed that supertunneling is possible for the boundary separating spin-1 fermions and Dirac charge carriers. The transmission coefficient (13) does not depend on the absolute value $k$, but has a significant angular dependence (see also [12]). Thus, the expression (13) cannot be associated with an effective barrier that would be characterized by any dimensional values (barrier height, effective thickness).

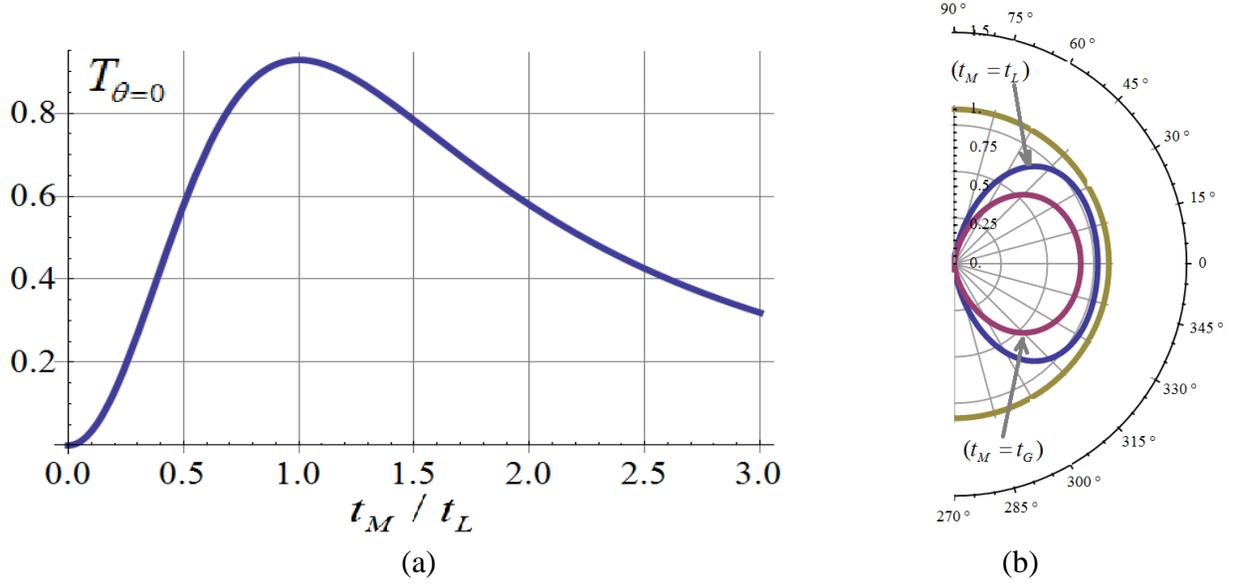

Fig. 2. Dependence of the transmission coefficient on the overlap integral at normal incidence (a) and its angular dependence (b).

**3.2.** For a macroscopic description of a conductor as a continuous medium, it is convenient to write boundary conditions in the form of a relationship between the wave functions (7) and (10) before and after the boundary. To exclude quantities that characterize the system at the microscopic level from boundary conditions, we write them for amplitudes of wave functions $\psi_{in,out}^{I,II,L}$,

$$\frac{\varphi^L(\vec{r})}{\sqrt{S_L}} = \psi_{in}^L \varphi_\theta^L e^{i(\vec{k}_C^L+\vec{k})\vec{r}} + \psi_{out}^L \varphi_{(\pi-\theta)}^L e^{i(\vec{k}_C^L+\vec{k}^*)\vec{r}},$$

$$\frac{\varphi^G(\vec{r})}{\sqrt{S_G}} = \psi_{in}^I \varphi_\theta^G e^{i(\vec{k}_C^I+\vec{k}^*)\vec{r}} + \psi_{in}^{II} \varphi_{(\pi-\theta)}^G e^{i(\vec{k}_C^{II}+\vec{k}^*)\vec{r}} + \psi_{out}^I \varphi_{(\pi-\theta)}^G e^{i(\vec{k}_C^I+\vec{k})\vec{r}} + \psi_{out}^{II} \varphi_\theta^G e^{i(\vec{k}_C^{II}+\vec{k})\vec{r}},$$

(14)

where $\vec{k}^* = (-k\sin\theta, k\cos\theta)$, and coefficients $\psi_{in,out}^{L,I,II}$ are chosen so that, in a macroscopic description, their squared modulus determines the physical density of charge carriers, in contrast to $\varphi^{L,G}$, which were defined as quantities whose squared modulus determines the total probability of the appearance of an electron at a given lattice site in the tight-binding approximation (1) in accordance with (7) and (10).

Then the tunneling of the barrier by a wave of arbitrary type can be written as a matrix that relates the wave function $\psi_{in} = (\psi_{in}^L, \psi_{in}^I, \psi_{in}^{II})^T$ of an electron moving toward the boundary from the Lieb lattice (L) or from graphene (I, II) and the wave function of an electron moving away from the boundary of the lattices, $\psi_{out} = \hat{U}\psi_{in}$. This matrix has the form:

$$\hat{U} = \begin{pmatrix} u_1 & u_3 \cos\frac{\theta}{2} & -u_3 \sin\frac{\theta}{2} \\ -u_3 \sin\frac{\theta}{2} & -\frac{1+u_2}{2}\sin\theta & \frac{(1+u_2)(1-\cos\theta)}{2} - 1 \\ u_3 \cos\frac{\theta}{2} & \frac{(1+u_2)(1+\cos\theta)}{2} - 1 & -\frac{1+u_2}{2}\sin\theta \end{pmatrix}, \quad (15)$$

where coefficients

$$u_1 = C\left(\cos^2\theta + i\cos\theta - \sqrt{3}\right), \quad u_3 = 2\sqrt[4]{3}\, C \cos\theta,$$
$$u_2 = C\left(-\cos^2\theta + i\cos\theta + \sqrt{3}\right), \quad C = \left(\cos^2\theta - i\cos\theta + \sqrt{3}\right)^{-1}, \quad (16)$$

describe boundary conditions in the subspace of wave functions $\psi_{in,out}^{L,O}$ orthogonal to the mirror mode (11) and (12),

$$\begin{pmatrix} \psi_{out}^L \\ \psi_{out}^O \end{pmatrix} = \begin{pmatrix} u_1 & u_3 \\ u_3 & u_2 \end{pmatrix} \begin{pmatrix} \psi_{in}^L \\ \psi_{in}^O \end{pmatrix}, \quad (17)$$

where expressions (15) and (17) are related by $\psi_{in} = (\psi_{in}^L, \psi_{in}^O \cos\frac{\theta}{2}, -\psi_{in}^O \sin\frac{\theta}{2})^T$, $\psi_{out} = (\psi_{out}^L, -\psi_{out}^O \sin\frac{\theta}{2}, \psi_{out}^O \cos\frac{\theta}{2})^T$. Considering the equality of the velocity $v = v_F$ for both lattices, the condition for conservation of the flux is reduced to the equality of the electron densities

$$\left|\psi_{in}^L\right|^2 + \left|\psi_{in}^I\right|^2 + \left|\psi_{in}^{II}\right|^2 = \left|\psi_{out}^L\right|^2 + \left|\psi_{out}^I\right|^2 + \left|\psi_{out}^{II}\right|^2, \quad (18)$$

which automatically follows from the unitarity of matrices (15) and (17).

**4. Edge states in a magnetic field.**

**4.1.** The behavior of the system shown in Fig. 1 is investigated in a quantizing magnetic field $\vec{B} = (0,0,B)$, the Coulomb gauge $\vec{A} = (0, Bx, 0)$ will be used. We assume that the radius of the Larmor orbit $r_L = cE/(|e|Bv_F)$ satisfies the inequality:

$$a_{L,G} \ll r_L \ll l_0, \qquad (19)$$

where $l_0$ is the mean free path, $e = -|e|$ is the electron charge, $B > 0$. For simplicity, we will not take into account the Zeeman splitting of energy levels $\Delta_B = g\mu_B B/2$, caused by the interaction of a normal spin with the magnetic field, $\Delta_B \ll E_{n+1} - E_n$, which is equivalent to the condition

$$B \ll \frac{1}{n} \frac{|e|\hbar v_F^2}{g^2 \mu_B^2 c}, \qquad (20)$$

here $\mu_B$ is the Bohr magneton, and g is an effective g-factor of the 2D system.

We restrict ourselves to a semiclassical approximation for wave functions on both sides of the boundary, after which we will use the matrix (15). The semiclassical description is convenient for physical interpretation, including allowing direct visibility of phase relationships, although the role of the Berry phase is clearly shown only in limiting cases, as will be shown below. The semiclassical wave function in graphene without taking into account the normalization factor has the form:

$$\varphi^I(\vec{r}) \sim \begin{pmatrix} c^-(x) \\ ic^+(x) \end{pmatrix} e^{i(\vec{k}_C^I \vec{r} + k_y y)}, \quad \varphi^{II}(\vec{r}) \sim \begin{pmatrix} c^+(x) \\ ic^-(x) \end{pmatrix} e^{i(\vec{k}_C^{II} \vec{r} + k_y y)}. \qquad (21)$$

Here

$$c^\pm(x) = \frac{1}{\sqrt{p^{(\pm)}(x)}} \cos\left\{\frac{1}{\hbar}\int_x^{x_0^\pm} p^{(\pm)}(\tilde{x})d\tilde{x} - \frac{\pi}{4}\right\},$$

$$p^{(\pm)}(x) = \sqrt{\frac{E^2}{v_F^2} \pm \frac{|e|B\hbar}{c} - \left(\frac{|e|B}{c}\right)^2 (x - X_c)^2}, \qquad (22)$$

where $X_c = -cP_y/(|e|B)$ is the centre of Larmor orbit and $x_0^\pm > 0$ is the root of the equation $p^{(\pm)}(x_0^\pm) = 0$.

Similarly for the Lieb lattice:

$$\varphi^L(\vec{r}) \sim \begin{pmatrix} \varphi_1(x) \\ \varphi_2(x) \\ \varphi_0(x) \end{pmatrix} e^{i(\vec{k}_C^L \vec{r} + k_y y)}. \qquad (23)$$

where $\varphi_0(x)$ is determined in the semiclassical approximation by the expression:

$$\varphi_0(x) \approx \frac{1}{\sqrt{p(x)}} \cos\left\{\frac{1}{\hbar}\int_{x_0^L}^{x} p_x(\tilde{x})d\tilde{x} - \frac{\pi}{4}\right\}, \tag{24}$$

where $x_0^L < 0$ is the root of the equation $p_x(x_{max}^\pm) = 0$,

$$p_x(x) = \sqrt{\frac{E^2}{v_F^2} - \left(\frac{|e|B}{c}\right)^2 (x - X_c)^2}, \tag{25}$$

which does not contain the additional phase shift caused by the Berry phase $\phi_B$, unlike (22). The other two components (23) can be obtained from the expression (24) by the relations that follow directly from (8):

$$\varphi_{1,2}(x) = -\frac{\hbar v_F}{\sqrt{2}E}\left(\frac{\partial}{\partial x} \mp ik_y \pm \frac{ix}{l_B^2}\right)\varphi_0(x) \tag{26}$$

where $\varphi_1(x)$ corresponds to the upper sign of the expression, and $l_B = \sqrt{c\hbar/(|e|B)}$ is the magnetic length.

According to (19), near the boundary, each of solutions (21) and (23) can be represented as the sum of two traveling waves, which can be written using coefficients $\psi_{in,out}^{L,I,II}$ (14). Then (21) and (23) lead to a relationship that can be conveniently written in the matrix form:

$$\psi_{in} = \hat{\Lambda}\psi_{out}, \quad \hat{\Lambda} = diag(e^{i\gamma_L}, e^{i\gamma_I}, e^{i\gamma_{II}}). \tag{27}$$

Expressions (21) - (26) lead to the following values of phase shifts:

$$\gamma_L = \frac{2}{\hbar}\int_{x_0^L}^{0} p_x dx - \frac{\pi}{2} = \frac{c}{|e|B\hbar}S_{p^2}^L - \frac{\pi}{2}, \quad \gamma_I = \gamma_{II} = \frac{2}{\hbar}\int_{0}^{x_0^G} p_x dx - \frac{\pi}{2} = \frac{c}{|e|B\hbar}S_{p^2}^G - \frac{\pi}{2}. \tag{28}$$

Here $x_0^G > 0$ ($x_0^L < 0$) is the positive (negative) root of equation $p_x(x) = 0$ (25). The phases (28) are determined by the areas of the segments $S_{p^2}^{L,G}$ of the orbit of a semiclassical electron with energy $E$ and centered at $X_c = -\frac{cp_F}{|e|B}tg\,\theta$, where $p_F = \frac{E}{v_F}$. The segment $S_{p^2}^L$ ($S_{p^2}^G$) corresponds to a part of the orbit in the Lieb lattice (in graphene),

$$S_{p^2}^{L,G} = \frac{p_F^2}{2}\left(\pi \pm 2\theta \pm \sin(2\theta)\right), \tag{29}$$

where the upper sign corresponds to $S_{p^2}^L$ and lower to $S_{p^2}^G$

An edge state exists if its wave function satisfies both equations $\psi_{out} = \hat{U}\psi_{in}$ (15) and $\psi_{in} = \hat{\Lambda}\psi_{out}$ (27), which is possible if

$$det(\hat{U}\hat{\Lambda} - \hat{I}) = 0 \qquad (30)$$

where $\hat{I}$ is a unit matrix. Equation (30) leads to the following expression, which determines the energy of quantum edge states through areas $\tilde{S}_{L,G} = \frac{c}{|e|B\hbar} S_{p^2}^{L,G}$:

$$\tilde{S}_L = 2\pi n - \frac{\pi}{2} + 2Arg\left[\sqrt{3}(\sin\theta + \sin\tilde{S}_G) + \cos\tilde{S}_G \cos\theta + \frac{2i}{\sqrt{3}}\frac{t_L t_G}{t_M^2}\cos\tilde{S}_G \cos^2\theta\right]. \qquad (31)$$

Expression (31) is an analog of the Bohr-Sommerfeld quantization rule and implicitly defines the energy spectrum of edge states $E_n = E_n(X_c)$, which is directly related to the total area $S_{p^2}^L + S_{p^2}^G = \pi p_F^2 = \pi E_n^2 / v_F^2$. The right side of (31) implicitly depends on $S_{p^2}^L$ through the angle $\theta$ (29). This leads to impossibility of any complete description of the edge states using simple analytical expressions. It is possible only for some special cases. Thus, the energy of edge states with the center $X_c = 0$ at the boundary of the lattices in the case $t_M = t_G$ can be written as

$$\begin{aligned} E_n^{(1)} &= v_F \sqrt{2\frac{|e|B\hbar}{c}\left(2n + \frac{3}{2}\right)}, \\ E_n^{(2)} &= v_F \sqrt{2\frac{|e|B\hbar}{c} 2n}, \\ E_n^{(3)} &= v_F \sqrt{2\frac{|e|B\hbar}{c}\left(2n + \frac{1}{2} + \frac{1}{\pi}arctg\frac{3 + 2\sqrt{3}}{2 + 2\sqrt{3}}\right)}, \end{aligned} \qquad (32)$$

where n is an integer.

**4.2.** Energy spectrum of edge states. Fig. 3.a shows the dependence of the energy of edge levels on the value $X_c$. This dependence can be obtained by a self-consistent numerical solution of equations (31) and (29) with the exception of the region near the edge of the graph, $0 < R_L - X_c \ll R_L$. The representation of the energy spectrum in the form (31) has a compact form, but its numerical solution by the iteration method converges poorly at $S_{p^2}^L \ll S_{p^2}^G$.

More convenient for numerical calculations is the method given in Appendix A, which is applicable in the entire range of where the WKB approximation is justified. In Fig. 3, the almost oscillatory nature of the dependence $E(X_c)$ is clearly visible, which can be represented as

$$S^L_{p^2} + 2 S^G_{p^2} = 2\pi \frac{eB\hbar}{c}(n+\nu), \qquad (33)$$

where $n$ is an integer, $\nu$ is a small oscillating correction. Far from the boundary $\left|S^L_{p^2} - S^L_{p^2}\right| \ll S^{L,G}_{p^2}$ and for $n \gg 1$, the behavior of the correction $\nu$ can be described by an almost periodic dependence $\nu(\tilde{S}_L, \tilde{S}_G) \sim \nu(\tilde{S}_L + 2\pi, \tilde{S}_G) \sim \nu(\tilde{S}_L, \tilde{S}_G + \pi)$, which is clearly seen in Fig.2.b.

For one period, one or three intersections with the Fermi level are possible, which can be easily shown by using substitution (A2) in the appendix, where the equation $E(X_c) = \varepsilon_F$ reduces to the cubic equation (A3), which can have one or three real roots.

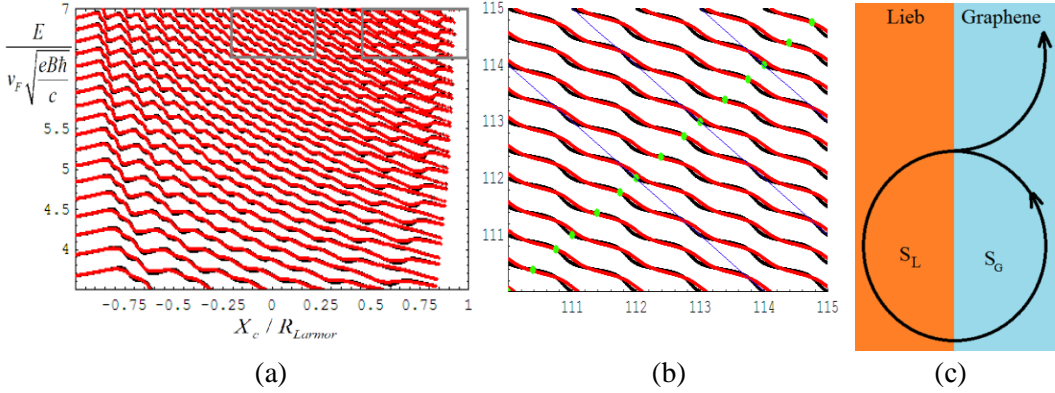

(a)  (b)  (c)

Fig. 3. Dependence of the energy of edge states $E$ on $X_c$ (a). Areas of segments according to the Bohr-Sommerfeld quantization rule (31) for the same edge states (b). The classical jumping orbit that can be associated with them (c).

The strong difference in areas $S^L_{p^2}$ и $S^G_{p^2}$ leads to a picture of edge states built along vertical and horizontal asymptotes (Fig. 4), which is explained by the small transmission coefficient $T(\theta) \ll 1$ (13) for <u>sliding</u> orbits. Indeed, in the case of zero transparency of the boundary, the magnetic quantization of conduction electrons should coincide with the independent quantization of areas $S^L_{p^2}$ and $S^G_{p^2}$, which would correspond to a simple system of vertical and horizontal lines in Fig. 4. Thus, we see a kind of transition process from the dependencies of Fig. 3.b to behavior of this kind. It can be noted that when moving from Fig. 4.a to Fig. 4.b, a phase shift $\pi$ is observed for the horizontal asymptote

$$S^G_{p_2} \approx 2\pi \frac{|e|B\hbar}{c}(n+\nu), \quad \nu|_{S^G_{p_2} \ll S^L_{p_2}} = \frac{3}{4}, \quad \nu|_{S^G_{p_2} \gg S^L_{p_2}} = \frac{1}{4}. \tag{34}$$

While the vertical asymptotes are described by the same expression

$$S^L_{p_2} \approx 2\pi \frac{|e|B\hbar}{c}(n+\frac{3}{4}) \tag{35}$$

in both extreme cases: $S^G_{p_2} \gg S^L_{p_2}$ (Fig. 4.a) and $S^G_{p_2} \ll S^L_{p_2}$ (Fig. 4.b). This behavior can be explained by the influence of the Berry phase, which is equal to $\pi$ present in graphene but absent in the Lieb lattice. Unfortunately, the question of the contribution of the Berry phase to parts of the trajectory is, strictly speaking, devoid of physical meaning, because the division of the phase into the contribution from the trajectory and the phase relations at the boundary is quite arbitrary, since it depends on the choice of the type of wave function. Only the total contribution remains constant, and this is what the measurable characteristics of the system are associated with. However, we note that the phase correction $\nu = \frac{3}{4}$ for $S^L_{p_2}$ (35) corresponds to a simple sum of the contributions from the turning point $\Delta\nu = \frac{1}{4}$ and reflection from the boundary $\Delta\nu = \frac{1}{2}$. While in graphene (34) $\nu = \frac{1}{4}$ coincides with the expected phase value after adding the Berry phase $\pi$ in the case $S^G_{p_2} \gg S^L_{p_2}$. Since when moving along the arc of the segment $S^G_{p_2}$ the direction of the electron velocity changes almost by $2\pi$. While for a small segment in the case $S^G_{p_2} \ll S^L_{p_2}$ the influence of the Berry phase can be neglected and $\nu = \frac{3}{4}$.

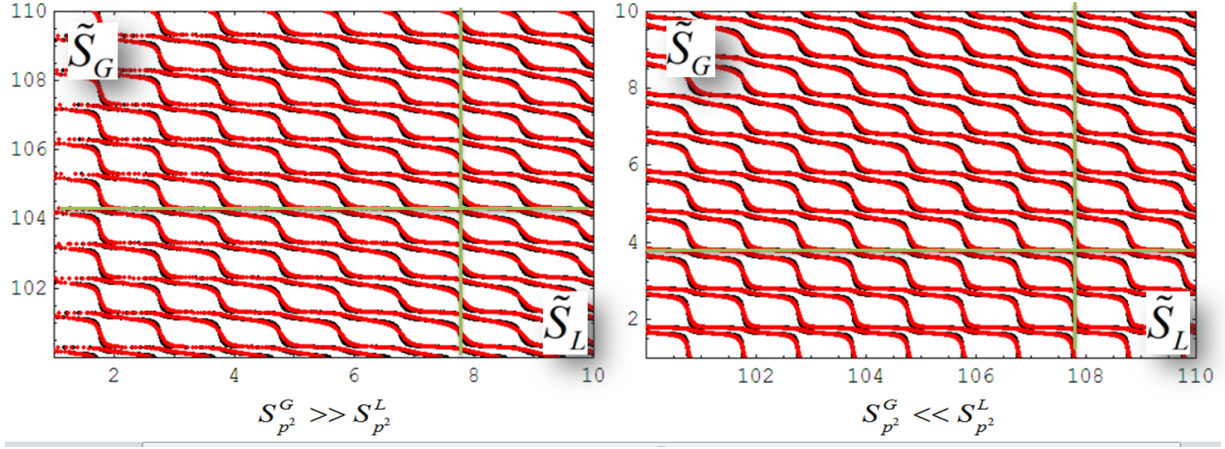

Fig. 4. Edge states for $t_M = t_L$ (red), $t_M = t_G$ (black) and their asymptotes (green) at $S^G_{p^2} \gg S^L_{p^2}$ (a) and $S^G_{p^2} \ll S^L_{p^2}$ (b).

**4.3.** The contribution to the density of states from the edge states can be written as:

$$\nu(E) = \sum_n \int \frac{dP_y}{2\pi\hbar} \delta(E - \epsilon_n(P_y)) = \frac{1}{2\pi\hbar} \sum_n \frac{1}{|v(P_n^y)|}, \quad v(P_n^y) = \frac{\partial E_n}{\partial P_y} = \frac{c}{|e|B} \frac{\partial E_n}{\partial X_c}, \quad (36)$$

where $\nu(E)$ is the density of edge states per unit length of the boundary, $E_n(X_c)$ determined by formula (A1). Expression (36) has a simple form, however the volume of expressions (A4-A6) does not give hope for writing the result of calculation (36) in a compact analytical form. Fig.5 shows the dependence of the density of edge states, calculated by the formula (36) at $t_M = t_G$.

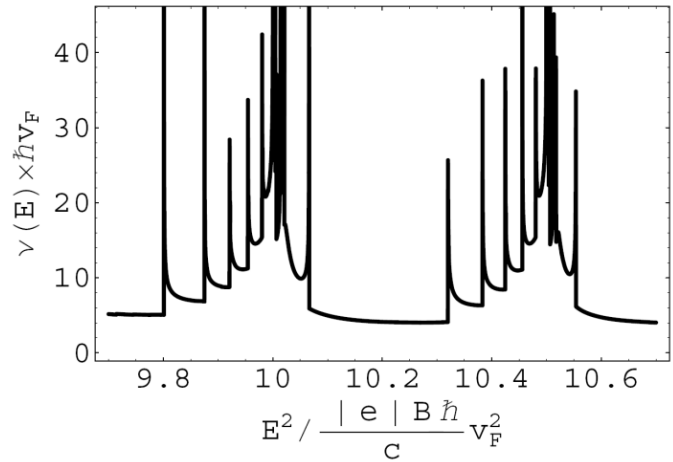

Fig.5. Density of edge states at $t_M = t_G$.

The figure shows that the density of edge states is characterized by a series of root singularities. These features arise in pairs due to a peculiar topological Lifshitz transition in branches of the one-dimensional energy spectrum of edge states $E_n(P_y)$. At intervals much larger than the de Broglie length, the energy of the edge states $E_n(X_c)$ decreases (Fig. 3.a). The monotonic trend of decreasing with increasing is explained by the fact that the Dirac spectrum in graphene is doubly degenerate due to the presence of two cones (according to the theorem on the paired number of cones [13]), unlike the Lieb lattice, which has a single cone point in the spectrum. In this way, the total number of states in graphene in the range $0 < E < E_0$ coincides with the number of states in the Lieb lattice in $0 < E < 2E_0$. The corresponding level density can be seen on the left and right edges of Fig.3.a, which ensures the agreement of the edge states with the simple Landau levels inside both lattices. Violation of the general trend $\frac{\partial E_n}{\partial X_c} < 0$ naturally leads to the appearance of a minimum and maximum in $E_n(X_c)$ and paired features in the density of states: $\nu \sim (E - E_{min})^{-1/2}$ and $\nu \sim (E_{max} - E)^{-1/2}$, which are clearly visible in Fig.5.

## 5. Conclusions

Edge states arising in the magnetic field at the interface of two conductors are described: with Dirac-type charge carriers and so-called "spin-1" fermions, which are realized in graphene-like and Lieb lattices, respectively. On the one hand, these states are a variety of long-known edge magnetic states [10]. On the other hand, the different topological nature of the two conductors forming the edge is revealed. It is worth emphasizing the special role of the so-called "No-go theorem" for chiral fermions [13], which speaks of the obligatory double degeneracy of the Dirac cone in graphene compared to the Lieb lattice, since the Berry phase along the contour of the entire reciprocal cell must be a multiple $2\pi$. Firstly, this leads to the obligatory suppression of the Fermi level by half of the discrete quantum levels from their number in graphene (Fig. 3.a). Secondly, this theorem indicates that the model used in the presented work is in a certain sense the simplest, since it uses the minimum allowed total number of cone points in the electron energy spectrum for materials on both sides of the boundary.

The characteristic values of the magnetic field, as well as other parameters characterizing edge states, will naturally depend on its specific experimental implementation, which can be quite exotic. First of all, because of the Lieb lattice, for which materials are used that are often very far from traditional conductors [6,8,14]. As an example only, we will give estimates if the conical energy spectrum is characterized by the value $v_F \approx 10^6 m/c$, as in graphene. Then smearing of peaks in Fig. 5 will be insignificant if the relaxation time $\tau$ satisfies the inequality $\frac{1}{\tau} \ll \frac{\Omega_1}{n}$, where $n$ – Landau level number, and $\Omega_1 = v_F\sqrt{2|e|B/(c\hbar)}$ -- the frequency of the transition between the zero and first Landau levels of the Dirac cone, $\Omega_1 \approx 5 \times 10^{13}$ Hz for $B = 1$T . The root dependence of the quantum levels of the Dirac cone $\epsilon_n \sim \sqrt{B}$ leads to the fact that at sufficiently small magnetic fields the Zeeman splitting $\Delta_B \sim B$ can be neglected. This is possible in the case of an insignificant restriction on the Landau level number and the magnetic field. It is possible in the case of an insignificant restriction for the Landau level number and magnetic field. Naturally, this estimate is limited to the case of ordinary Dirac conductors and will not be valid for systems in which the real spin participates in the formation of the cone spectrum.

**Appendix**

The equation for the coordinate $X_c$ is the following from (31),

$$\epsilon_n(\theta, X_c) = E, \tag{A1}$$

can be solved through the roots of a cubic equation by using the change of variables

$$z = tg(\frac{\tilde{S}_G - \tilde{S}_L}{4}), \quad \tilde{S}_c = \tilde{S}_L + \tilde{S}_G, \tag{A2}$$

where $\frac{|e|B\hbar}{c}\tilde{S}_c(E) = \pi p_F^2 = \pi E^2/v_F^2$ is the area of the sum of two segments. After simple transformations, the expression (31) can be rewritten as:

$$az^3 + bz^2 + cz + d = 0, \tag{A3}$$

where

$$a = \sqrt{3}(\sin\theta - \sin\frac{\tilde{S}_c}{2}) - \cos\theta\cos\frac{\tilde{S}_c}{2} - \frac{2}{\sqrt{3}}\frac{t_L t_G}{t_M^2} tg\left(\frac{\tilde{S}_c}{4} + \frac{\pi}{4}\right)\cos^2\theta\cos\frac{\tilde{S}_c}{2}, \tag{A4}$$

$$b = \sqrt{3}tg\left(\frac{\tilde{S}_c}{4}+\frac{\pi}{4}\right)(\sin\frac{\tilde{S}_c}{2}-\sin\theta)-2\cos\theta\sin\frac{\tilde{S}_c}{2}+\cos\theta\cos\frac{\tilde{S}_c}{2}tg\left(\frac{\tilde{S}_c}{4}+\frac{\pi}{4}\right)$$

$$-\frac{2}{\sqrt{3}}\frac{t_L t_G}{t_M^2}\cos^2(\theta)\left(\cos(\frac{\tilde{S}_c}{2})+2\sin(\frac{\tilde{S}_c}{2})tg\left(\frac{\tilde{S}_c}{4}+\frac{\pi}{4}\right)\right)+2\sqrt{3}\cos(\frac{\tilde{S}_c}{2}),$$

$$c = \sqrt{3}(\sin\theta+\sin\frac{\tilde{S}_c}{2})+\cos\theta\cos\frac{\tilde{S}_c}{2}+tg\left(\frac{\tilde{S}_c}{4}+\frac{\pi}{4}\right)\left(2\cos\theta\sin\frac{\tilde{S}_c}{2}-2\sqrt{3}\cos\frac{\tilde{S}_c}{2}\right)$$

$$+\frac{2}{\sqrt{3}}\frac{t_L t_G}{t_M^2}\cos^2\theta\left(\cos\frac{\tilde{S}_c}{2}tg\left(\frac{\tilde{S}_c}{4}+\frac{\pi}{4}\right)-2\sin\frac{\tilde{S}_c}{2}\right),$$

(A5)

$$d = -tg\left(\frac{\tilde{S}_c}{4}+\frac{\pi}{4}\right)\left(\sqrt{3}(\sin\theta+\sin\frac{\tilde{S}_c}{2})+\cos\theta\cos\frac{\tilde{S}_c}{2}\right)+\frac{2}{\sqrt{3}}\frac{t_L t_G}{t_M^2}\cos^2\theta\cos\frac{\tilde{S}_c}{2}. \qquad \text{(A6)}$$